\documentclass[aps,showpacs,twocolumn,superscriptaddress,pre]{revtex4}
\usepackage{amsmath}
\usepackage{amssymb}
\usepackage{mathtext}

\begin{document}

\newcommand {\e} {\varepsilon}
\newcommand {\ph} {\varphi}
\newcommand {\la} {\left\langle}
\newcommand {\ra} {\right\rangle}

\title{Large-scale thermal convection in a horizontal porous layer}

\author{Denis S.\ Goldobin}
\affiliation{Department of Theoretical Physics, Perm State University,
        15 Bukireva str., 614990, Perm, Russia}
\affiliation{Department of Physics and Astronomy, University of Potsdam,
        Postfach 601553, D--14415 Potsdam, Germany}
\author{Elizaveta V.\ Shklyaeva}
\affiliation{Department of Theoretical Physics, Perm State University,
        15 Bukireva str., 614990, Perm, Russia}

\begin{abstract}
In a range of physical systems the first instability in
Rayleigh-B\'ernard convection between nearly thermally insulating
horizontal plates is large-scale. This holds for thermal
convection of fluids saturating porous media. Large-scale thermal
convection in a horizontal layer is governed by remarkably similar
equations both in the presence of a porous matrix and without it,
with only one additional term for the latter case, which, however,
vanishes under certain conditions ({\it e.g.}, two-dimensional
flows or infinite Prandtl number).
We provide a rigorous derivation of long-wavelength equations for
a porous layer with inhomogeneous heating and possible pumping.
\end{abstract}

\maketitle

Rayleigh-B\'ernard convection, {\it i.e.}, a thermal buoyancy
convection in a layer confined between two rigid horizontal plates
and uniformly heated from below, is one of the classical problems
of fluid dynamics. When the bounding plates are nearly thermally
insulating, the first instability of the heat conductive state
(quiescent state) of the layer is long-wavelength
(large-scale)~\cite{Sparrow_etal-1964,Ingham-Pop-1998,porous_media,Goldobin-Lyubimov-2007},
{\it i.e.}, horizontal gradients of the fluid velocity field are
small compared to vertical ones and, thus, the horizontal scale of
the flow is large compared to the layer height. For large-scale
patterns, temperature perturbations $\theta$ are nearly uniform
along the vertical coordinate $z$ and in terms of $\theta(x,y)$
the problem is two-dimensional (2D). Making allowance for the
heating inhomogeneity, one can find
\begin{eqnarray}
&\hspace{-3mm}
\partial_t\theta\!+\!\Delta^2\theta
-\!\nabla\!\cdot\!(\theta\vec{s})
+\!\nabla\!\cdot\!\left[q(x,y)\nabla\theta-\!\nabla\theta|\nabla\theta|^2\right]=0,&
\label{lstc-01}\\
&\hspace{-3mm}
\Delta\vec{s}=0\,,\quad [\nabla\times\vec{s}]_z\equiv B
 =\frac{3}{2Pr}[\nabla\times(\nabla\theta\,\Delta\theta)]_z.&
\label{lstc-02}
\end{eqnarray}
Here $\partial_t$ denotes the partial time derivative, $q(x,y)$ is
the relative departure of the heat flux (imposed by the heating)
through the boundaries from the critical value for a homogeneous
heating, Prandtl number $Pr\!=\!\nu/\chi$ is the ratio of the
kinematic viscosity and the heat diffusivity, $\vec{s}$ is
responsible for advectional heat transfer.
Knobloch~\cite{Knobloch-1990} has discussed pattern selection for
Eq.\,(\ref{lstc-01}) with uniform $q$ and $\vec{s}\!=\!0$, which
is relevant, {\it e.g.}, to $Pr\!\to\!\infty$ or 2D setups where
flows are homogeneous in the $y$-direction. Later on, the term
missing in~\cite{Knobloch-1990} was accounted for
in~\cite{Shtilman-Sivashinsky-1991}. Remarkably,
in~\cite{Aristov-Frick-1989}, equations similar to~(\ref{lstc-01})
have been obtained for large-scale turbulence in
Rayleigh-B\'ernard convection.

Various problems related to large-scale thermal convection in
porous media have been repeatedly addressed in the literature
({\it e.g.},
\cite{porous_media,Goldobin-Lyubimov-2007}).
However, researches typically either deal with the case of a
rectangular cavity with a large aspect ratio (\cite{porous_media},
where no equations similar to~(\ref{lstc-01},\ref{lstc-02})
have been derived) or discuss a finite departure from the
stability threshold (\cite{Goldobin-Lyubimov-2007}; such a
departure leads to discontinuities of the velocity field, which
cannot be treated within the framework of the long-wavelength
approximation). Pattern formation under heating inhomogeneity has
been rather extensively studied for thin films
 ({\it e.g.}, \cite{Yeo-Craster-Matar-2003} and Refs.\ therein)
owing to important technical applications (rupture of lubricating
films~\cite{Warner-Craster-Matar-2002}, {\it etc.}). To the
authors' knowledge, for porous media the equation system similar
to~(\ref{lstc-01},\ref{lstc-02}) has not been presented in the
literature.

In this Brief Report, we introduce the specific physical system we
deal with. Then we derive the equation of large-scale convection
for this system and find that it is similar to
Eqs.\,(\ref{lstc-01},\ref{lstc-02}) with $B=0$, though the
relationships between the flow and the temperature perturbation
are different for different fluid dynamical
systems~\cite{Shtilman-Sivashinsky-1991,Aristov-Frick-1989,Goldobin-Lyubimov-2007}.

\paragraph{Thermal convection in a horizontal porous layer.}
Let us consider convection of a fluid saturating a horizontal
porous layer heated from below. Boundaries are impermeable, the
heat flux across the layer is fixed (implying the heat diffusivity
of the boundaries is small compared to the one of the porous
matrix saturated with the fluid), but inhomogeneous along the
layer. Relaxation of the local temperature difference between the
porous matrix and the fluid is assumed to be fast, and we do not
introduce several temperatures for them. For small perturbations
of the temperature about $T_0$, one may guess a linear dependence
of the fluid density on the temperature;
$\rho(T)=\rho_0(1-\beta(T-T_0))$, where $\rho_0=\rho(T_0)$,
$\beta=(\partial\rho/\partial T)_p$. The reference frame is such
that the $(x,y)$-plane is horizontal, $z=0$ and $z=h$ are the
lower and upper layer boundaries, respectively. We adopt the
conventional Darcy--Boussinesq
approximation~\cite{Ingham-Pop-1998},
\begin{eqnarray}
&0=-\rho_0^{-1}\nabla p-\nu K^{-1}\vec{v}
 +g\beta T\vec{e}_z\,,&\label{DB-1}\\
&\partial_tT
 +b^{-1}\nabla\cdot(\vec{v}T)=\chi\Delta T\,,&\label{DB-2}\\
&\nabla\cdot\vec{v}=0\,,&\label{DB-3}\\
&z=0,\,h\;:\quad v_z=0,\quad
 \partial_zT=-A(1+q(x,y))\,,&\label{DB-4}
\end{eqnarray}
where $\vec{v}$: the macroscopic filtration velocity, $K$: the
permeability, $\vec{g}=-g\vec{e}_z$: the gravity, $b$: the
specific heat capacity of the saturated porous medium divided by
the one of the fluid, $\chi$: the heat diffusivity of the
saturated porous medium, $\chi C_pA(1+q(x,y))$: the imposed heat
flux ($C_p$ is the specific heat capacity).

It is convenient to measure the length by layer height $h$, time
by $h^2/\chi$, the velocity by $b\chi/h$, the temperature by $Ah$,
and the pressure by $b\rho_0\nu\chi/K$. The dimensionless
parameter governing the behavior of the system is the
Rayleigh--Darcy number $R=\beta Ah^2gK/b\nu\chi$.

The dimensionless form of system\,(\ref{DB-1}--\ref{DB-4}) reads
\begin{eqnarray}
& -\nabla p-\vec{v}+R\, T\vec{e}_z=0\,,&\label{lstc-dl01}\\
& \partial_tT+\nabla\cdot(\vec{v}T)=\Delta T\,,&\label{lstc-dl02}\\
& \nabla\cdot\vec{v}=0\,,&\label{lstc-dl03}\\
& z=0,\,1\;:\quad v_z=0,\quad
\partial_zT=-1-q(x,y)\,.&\label{lstc-dl04}
\end{eqnarray}

\paragraph{Long-wavelength approximation.}
For a uniform fixed heat flux the first instability is known to be
long-wavelength~\cite{Ingham-Pop-1998}. This holds for $q(x,y)$
slowly varying in the boundary plane: $|\nabla q|/|q|\sim\e\ll1$.
In order to avoid large temperature gradients, which correspond to
jumps of derivatives of fields in the long-wavelength
limit~\cite{Goldobin-Lyubimov-2007}, we restrict ourselves to a
small supercriticality. Below we will find the critical value
$R_\mathrm{C}$ for a homogeneous heating and set $R=R_\mathrm{C}$;
hence, nearcritical regimes occur for small $q(x,y)$ which may be
thought to measure the relative departure of the Rayleigh--Darcy
number from the critical value.

Owing to Eq.\,(\ref{lstc-dl03}) the horizontal component of
velocity $\vec{v}$ is large compared to the vertical one because
variations of velocity $\vec{v}$ by shifts along the horizontal
directions are small compared to ones by shifts transversal to the
layer. Let us explicitly account for this fact,
 $\vec{v}=w\vec{e}_z+\e^{-1}\vec{u}$,
where $\vec{u}$ is the horizontal component of the filtration
velocity field, $w$ is the vertical one. Rescaling horizontal
coordinates, $(x,y)\to(\e^{-1}x,\e^{-1}y)$, explicitly writing
$q(x,y)=\e^2q_2(x,y)$, and projecting momentum conservation
law~(\ref{lstc-dl01}) onto the vertical and horizontal directions,
one may rewrite equation system~(\ref{lstc-dl01}--\ref{lstc-dl04})
in a form convenient for the further treatment;
\begin{eqnarray}
&\displaystyle -\partial_z p-w+R\,T=0\,,&\label{lstc-fdl01}\\
&\displaystyle \vec{u}=-\e^2\nabla_{\!2}p\,,&\label{lstc-fdl02}\\
&\displaystyle \partial_tT+\partial_z(wT)+\nabla_{\!2}\cdot(\vec{u}T)=\partial_z^2T+\e^2\Delta_2T\,,&\label{lstc-fdl03}\\
&\displaystyle \partial_zw+\nabla_{\!2}\cdot\vec{u}=0\,,&\label{lstc-fdl04}\\
&\displaystyle z=0,\,1\;:\quad w=0,\quad
\partial_zT=-1-\e^2q_2(x,y)\,.&\label{lstc-fdl05}
\end{eqnarray}
Here the subscript ``2'' for spatial derivatives means the
differentiation with respect to two horizontal coordinates.

Since $\e$ appears squared
in~(\ref{lstc-fdl01}--\ref{lstc-fdl05}), only even powers of $\e$
are present in the expansion: $w=w_0+\e^2w_2+\e^4w_4+\dots$,
$T=T_0+\e^2T_2+\e^4T_4+\dots$, {\it etc}. The long-wavelength
approximation assumes a {\it weak} spatial inhomogeneity of
temperature perturbations which results in a {\it slow} temporal
evolution. With only even powers of $\e$ in the expansion, one
expects characteristic times of the evolution of large-scale
patterns to be not lesser than $\propto\e^{-2}$; therefore, in
terms of ``slow'' times,
 $\partial_t=\e^2\partial_{t_2}+\e^4\partial_{t_4}+\dots$.


\noindent
\underline{$\e^0$}:\quad
In the leading order, problem~(\ref{lstc-fdl01}--\ref{lstc-fdl05})
yields
\begin{eqnarray}
& -\partial_z p_0-w_0+R\,T_0=0\,,&\label{lstc-001}\\
& \vec{u}_0=0\,,&\label{lstc-002}\\
& \partial_z(w_0T_0)+\nabla_{\!2}\cdot(\vec{u}_0T_0)=\partial_z^2T_0\,,&\label{lstc-003}\\
& \partial_zw_0+\nabla_{\!2}\cdot\vec{u}_0=0\,,&\label{lstc-004}\\
& z=0,\,1\;:\quad w_0=0,\quad
\partial_zT_0=-1\,.&\label{lstc-005}
\end{eqnarray}
From Eqs.\,(\ref{lstc-004},\ref{lstc-002}),\quad
$\partial_zw_0=0$, $w_0=C_1(x,y)=0$
\\[0pt]
[$C_1=0$ due to the boundary conditions (BCs)].

\noindent
From Eq.\,(\ref{lstc-003}),\quad
$\partial_z^2T_0=0$, {\it i.e.}, $T_0=C_2(x,y)z+\theta(x,y)$;
accounting for BCs~(\ref{lstc-005}), one obtains
\begin{equation}
T_0=-z+\theta(x,y)\,,
\label{lstc-006}
\end{equation}
where $\theta(x,y)$ should be determined from higher orders of the
expansion while here it appears as an unknown function of the
horizontal coordinates.

\noindent
From Eq.\,(\ref{lstc-001}),\quad
$\partial_zp_0=R\,T_0=-R\,z+R\,\theta(x,y)$,
\begin{equation}
\textstyle p_0=-\frac{1}{2}R\,z^2+R\,\theta(x,y)\,z+\varPi_0(x,y)\,,
\label{lstc-007}
\end{equation}
where $\varPi_0(x,y)$ is unknown in this order of the expansion.


\noindent
\underline{$\e^2$}:
\vspace{-7mm}
\begin{eqnarray}
& -\partial_zp_2-w_2+R\,T_2=0\,,&\label{lstc-201}\\
& \vec{u}_2=-\nabla_{\!2}p_0\,,&\label{lstc-202}\\
& \partial_{t_2}T_0+\partial_z(w_2T_0)+\nabla_{\!2}\!\cdot\!(\vec{u}_2T_0)=\partial_z^2T_2+\Delta_2T_0\,,&\label{lstc-203}\\
& \partial_zw_2+\nabla_{\!2}\cdot\vec{u}_2=0\,,&\label{lstc-204}\\
& z=0,\,1\;:\quad w_2=0,\quad
\partial_zT_2=-q_2(x,y)\,.&\label{lstc-205}
\end{eqnarray}
From Eq.\,(\ref{lstc-202}),
\[
\textstyle\vec{u}_2=-\nabla_{\!2}p_0=-R\,z\nabla_{\!2}\theta(x,y)-\nabla_{\!2}\varPi_0(x,y).
\]
From Eq.\,(\ref{lstc-204}),
\begin{eqnarray}
&\textstyle
\partial_zw_2=-\nabla_{\!2}\cdot\vec{u}_2=R\,z\Delta_2\theta(x,y)+\Delta_2\varPi_0(x,y),&\nonumber\\
&\textstyle
w_2=\frac{1}{2}R\,z^2\Delta_2\theta(x,y)+\Delta_2\varPi_0(x,y)z+C_3(x,y).&\nonumber
\end{eqnarray}
BCs~(\ref{lstc-205}) for the velocity field yield
\begin{eqnarray}
&\textstyle z=0:&\textstyle
 C_3=0\,,\nonumber\\
&\textstyle z=1:&\textstyle
 \Delta_2\varPi_0(x,y)=-\frac{1}{2}R\,\Delta_2\theta(x,y)\,.\nonumber
\end{eqnarray}
From the latter BC,
\[
\textstyle
\varPi_0(x,y)=-\frac{1}{2}R\,\theta(x,y)+\pi_0(x,y)\,,\quad\Delta_2\pi_0(x,y)=0\,.
\]

Note, $\la\vec{u}_2\ra=-\nabla_{\!2}\pi_0$ (henceforth, $\la
f\ra\equiv\int_0^1f\,\mathrm{d}z$). Let us consider a layer domain
limited in the horizontal directions by boundary $\varGamma$ (to
keep it simple, we assume $\varGamma$ to be vertical), where a
fixed fluid gross flux (or its absence; ``gross'' means averaged
over $z$, and, in particular, the absence of the gross flux does
not necessarily claim the absence of the flux) is imposed.
Mathematically, this means that the orthogonal to $\varGamma$
component of $\la\vec{u}_2\ra$ is fixed: $\la\vec{u}_2\ra_n=Q$
($Q$ is not uniform along $\varGamma$; owing to mass conservation,
 $\int_\varGamma Q\,\mathrm{d}\varGamma=0$).
Hence, one obtains the boundary problem for $\pi_0(x,y)$,
\begin{equation}
\Delta_2\pi_0=0,\qquad
\left.\partial_n\pi_0\right|_\varGamma\!=-Q\quad
\textstyle
(\,\int_\varGamma Q\,\mathrm{d}\varGamma=0\,),
\label{lstc-206}
\end{equation}
where $\partial_n$ is the orthogonal to $\varGamma$ component of
the gradient. This problem has a unique solution (up to an
insignificant constant) unambiguously defined by $Q$. Thus,
$\pi_0(x,y)$ describes an imposed advection (pumping) in the
layer, which is caused and {\it unambiguously} controlled by BCs
on $\varGamma$ (the pressure or the fixed gross flux). As soon as
we consider a nearcritical behavior, it makes sense to not allow
for an imposed advection in the leading order, otherwise a
moderately strong advection would overpress the effect of a weakly
inhomogeneous heating. We set
\begin{equation}
\pi_0(x,y)=0
\label{lstc-207}
\end{equation}
and will take the imposed advection into account in higher orders.

With (\ref{lstc-207}), one can write down
\begin{eqnarray}
&\vec{u}_2=-R\left(z-\frac{1}{2}\right)\nabla_{\!2}\theta(x,y)\,,&\label{lstc-208}\\
&w_2=\frac{1}{2}R\,(z^2-z)\,\Delta_2\theta(x,y)\,.&\label{lstc-209}
\end{eqnarray}

Let us now integrate~(\ref{lstc-203}) over $z\in[0,1]$;
\begin{equation}
\partial_{t_2}\!\la T_0\ra+\la\partial_z(w_2T_0)\ra+\nabla_{\!2}\!\cdot\!\la\vec{u}_2T_0\ra
 =\la\partial_z^2T_2\ra+\Delta_2\la T_0\ra.
\label{lstc-210}
\end{equation}
Due to BCs~(\ref{lstc-205}),
 $\int_0^1\partial_z(w_2T_0)\,\mathrm{d}z=w_2T_0|_{z=0}^1=0$,
 $\int_0^1\partial_z^2T_2\,\mathrm{d}z=\partial_zT_2|_{z=0}^1=-q_2(x,y)|_{z=0}^1=0$.
Whereas,
\[
\nabla_{\!2}\cdot\la\vec{u}_2T_0\ra=-\nabla_{\!2}\cdot\la\vec{u}_2z\ra+\nabla_{\!2}\cdot(\la\vec{u}_2\ra\theta)\,.
\]
Substituting (\ref{lstc-208}) into the first term of the rhs of
the last equation and accounting for
$\la\vec{u}_2\ra=-\nabla_{\!2}\pi_0=0$, one finds
\[
\nabla_{\!2}\cdot\la\vec{u}_2T_0\ra
 =R\,\Delta_2\theta\la z^2-z/2\ra
 =(R/12)\Delta_2\theta\,.
\]
For the rest of the terms in Eq.\,(\ref{lstc-210}),
$\partial_{t_2}T_0=\partial_{t_2}\theta$,
$\Delta_2T_0=\Delta_2\theta$, and Eq.\,(\ref{lstc-210}) finally
reads
\begin{equation}
\partial_{t_2}\theta=(1-R/12)\Delta_2\theta\,.
\label{lstc-211}
\end{equation}

For $R<12$, Eq.\,(\ref{lstc-211}) is a conventional diffusion
equation, and for trivial BCs (on $\varGamma$) or an infinite
layer, all inhomogeneities of $\theta$ decay. For $R>12$, it is a
diffusion equation with a negative diffusivity, where all the
inhomogeneous perturbations grow. Thus, $R=12$ is the linear
stability threshold of the system. Note, nonlinearity does not
play a role in this order of the expansion. In order to account
for nonlinear effects and the dependence of the linear stability
on the wavelength (now all the perturbations either grow or decay
regardless to their wavelength), we should restrict ourselves to
the vicinity of the stability threshold. For this purpose, we set
\[
R=R_\mathrm{C}=12
\]
and introduce departure from the threshold via $q$. Since the
local Rayleigh--Darcy number (``local'' means defined for a small
domain of the layer) $R_\mathrm{local}=R(1+q)$, positive $q$
corresponds to a supercritical regime, negative $q$ does to a
subcritical one. As soon as $R=12$, $\partial_{t_2}\theta=0$;
therefore, we should consider a slower evolution,
$\partial_t\theta=\e^4\partial_{t_4}\theta+\dots$.

Let us now derive $T_2$ from Eq.\,(\ref{lstc-203}).
\begin{eqnarray}
&&\;
\partial_z^2T_2=\partial_z(w_2T_0)+\nabla_{\!2}\!\cdot\!(\vec{u}_2T_0)-\Delta_2T_0\,;
\nonumber\\[5pt]
&&\hspace{-4mm}\textstyle
 T_2=-\left(\frac{3}{2}z^4-2z^3\right)\Delta_2\theta
 +(2z^3-3z^2)\theta\Delta_2\theta
\nonumber\\
&&\textstyle
 +(z^4-z^3)\Delta_2\theta-(2z^3-3z^2)\nabla_{\!2}\!\cdot\!(\theta\nabla_{\!2}\theta)-\frac{1}{2}z^2\Delta_2\theta
\nonumber\\
&&\textstyle
 +C_4(x,y)\,z+\theta_2(x,y)\,.
\nonumber
\end{eqnarray}
Due to the relation $\nabla_{\!2}\cdot(\theta\nabla_{\!2}\theta)
=\theta\Delta_2\theta+(\nabla_{\!2}\theta)^2$, one can eliminate
the term $\theta\Delta_2\theta$ from the expression for $T_2$;
\begin{eqnarray}
&&\hspace{-5mm}\textstyle
 T_2=\left(-\frac{1}{2}z^4+z^3-\frac{1}{2}z^2\right)\Delta_2\theta
\nonumber\\
&&\textstyle
 +(-2z^3+3z^2)(\nabla_{\!2}\theta)^2+C_4(x,y)z+\theta_2(x,y).
\label{lstc-212}
\end{eqnarray}
Owing to BCs~(\ref{lstc-205}), $C_4(x,y)=-q_2(x,y)$;
$\theta_2(x,y)$ is still undefined.

Remarkably, $\theta$ and $\theta_2$ depend on $z$ in one and the
same fashion ({\it i.e.}, are uniform along $z$) and are for the
moment undetermined functions of $x$ and $y$. Hence, $\theta_2$
may be chosen as one needs, and this will be automatically
balanced by $\theta$. Let us use this fact. Notice,
\[
\textstyle
\la T\ra=-\frac{1}{2}+\theta(x,y)+\e^2\la T_2\ra+O(\e^4)\,;
\]
therefore, if one defines $\theta_2$ in such a way as to make
 $\la T_2\ra=0$,
then $\theta$ will be a $z$-mean temperature up to the truncation
accuracy of our expansion, {\it i.e.}, $\e^4$\!. Thus,
\begin{eqnarray}
&\textstyle
\la T_2\ra=-\frac{1}{60}\Delta_2\theta+\frac{1}{2}(\nabla_{\!2}\theta)^2
 -\frac{1}{2}q_2(x,y)+\theta_2(x,y)=0\,,
 \nonumber\\[3pt]
&\textstyle
\theta_2=\frac{1}{60}\Delta_2\theta
 -\frac{1}{2}(\nabla_{\!2}\theta)^2+\frac{1}{2}q_2\,.
\label{lstc-213}
\end{eqnarray}

Finally,
\begin{eqnarray}
&&\hspace{-7mm}\textstyle
 T_2=\left(-\frac{1}{2}z^4+z^3-\frac{1}{2}z^2+\frac{1}{60}\right)\Delta_2\theta
\nonumber\\
&&\textstyle
 +\left(-2z^3+3z^2-\frac{1}{2}\right)(\nabla_{\!2}\theta)^2+\left(-z+\frac{1}{2}\right)q_2.
\label{lstc-214}
\end{eqnarray}

From Eq.\,(\ref{lstc-201}) follows $\partial_zp_2=-w_2+R\,T_2$,
and, integrating it with respect to $z$, one finds the pressure
\begin{eqnarray}
&&\hspace{-7mm}\textstyle
 p_2=\left(-\frac{6}{5}z^5+3z^4-4z^3+3z^2+\frac{1}{5}z\right)\Delta_2\theta
\nonumber\\
&&\textstyle
 +\left(-6z^4+12z^3-6z\right)(\nabla_{\!2}\theta)^2
\nonumber\\
&&\textstyle
 +\left(-6z^2+6z\right)q_2+\varPi_2(x,y)\,.
\label{lstc-215}
\end{eqnarray}


\noindent
\underline{$\e^4$}:\quad
As we will not construct the expansion beyond this order, we do
not have to consider all the equations. The following is
sufficient;
\begin{eqnarray}
& \vec{u}_4=-\nabla_{\!2}p_2\,,&\label{lstc-401}\\
&\hspace{-15mm}
 \partial_{t_4}T_0+\partial_z(w_4T_0)+\nabla_{\!2}\!\cdot\!(\vec{u}_4T_0)&\nonumber\\
&\hspace{10mm}
 +\partial_z(w_2T_2)+\nabla_{\!2}\!\cdot\!(\vec{u}_2T_2)
 =\partial_z^2T_4+\Delta_2T_2\,,&\label{lstc-402}\\
& \partial_zw_4+\nabla_{\!2}\cdot\vec{u}_4=0\,,&\label{lstc-403}\\
& z=0,\,1\;:\quad w_4=0,\quad
\partial_zT_4=0\,.&\label{lstc-404}
\end{eqnarray}

Without calculating $\vec{u}_4$, we may substitute
(\ref{lstc-401}) directly into (\ref{lstc-403});
$\partial_zw_4=\Delta_2p_2$. Then
\begin{eqnarray}
&&\hspace{-5mm}\textstyle
 w_4=\left(-\frac{1}{5}z^6+\frac{3}{5}z^5-z^4+z^3+\frac{1}{10}z^2\right)\Delta_2^2\theta
\nonumber\\
&&\textstyle
 +\left(-\frac{6}{5}z^5+3z^4-3z^2\right)\Delta_2(\nabla_{\!2}\theta)^2
\nonumber\\
&&\textstyle
 +\left(-2z^3+3z^2\right)\Delta_2q_2+z\Delta_2\varPi_2\,.
\label{lstc-405}
\end{eqnarray}
From BCs\,(\ref{lstc-404}),
\[
\textstyle
w_4|_{z=1}=0=\frac{1}{2}\Delta_2^2\theta
-\frac{6}{5}\Delta_2(\nabla_{\!2}\theta)^2+\Delta_2q_2+\Delta_2\varPi_2\,;
\]
therefore,
\[
\textstyle
\varPi_2=-\frac{1}{2}\Delta_2\theta
+\frac{6}{5}(\nabla_{\!2}\theta)^2-q_2+\pi_2\,,
\]
where $\Delta_2\pi_2=0$.

Notice, making use of Eq.\,(\ref{lstc-215}), one can calculate
\[
\la\vec{u}_4\ra=-\nabla_{\!2}\la p_2\ra=-\nabla_{\!2}\pi_2(x,y)\,.
\]
We have already established the relationship between
$\la\vec{u}_2\ra$ and $\pi_0$ and found $\pi_0$ to describe an
imposed advection. We have set $\pi_0=0$, but now it makes sense
to keep $\pi_2$, since it describes an advection imposed by
lateral boundary conditions.

Substituting $\varPi_2$ into Eq.\,(\ref{lstc-405}), one obtains
the final expression
\begin{eqnarray}
&&\hspace{-7mm}\textstyle
 w_4=\left(-\frac{1}{5}z^6+\frac{3}{5}z^5-z^4+z^3+\frac{1}{10}z^2-\frac{1}{2}z\right)\Delta_2^2\theta
\nonumber\\
&&\textstyle
 +\left(-\frac{6}{5}z^5+3z^4-3z^2+\frac{6}{5}z\right)\Delta_2(\nabla_{\!2}\theta)^2
\nonumber\\
&&\textstyle\qquad
 +\left(-2z^3+3z^2-z\right)\Delta_2q_2\,.
\label{lstc-406}
\end{eqnarray}

Now the integration of Eq.\,(\ref{lstc-402}) over $z\in[0,1]$
yields the evolution equation for $\theta$
\begin{eqnarray}
&&\hspace{-7mm}\textstyle
 \partial_{t_4}\theta+\la\partial_z(w_4T_0)\ra+\nabla_{\!2}\!\cdot\!\la\vec{u}_4T_0\ra
\nonumber\\
&&\textstyle
 +\la\partial_z(w_2T_2)\ra+\nabla_{\!2}\!\cdot\!\la\vec{u}_2T_2\ra
 =\la\partial_z^2T_4\ra+\Delta_2\la T_2\ra.
\nonumber
\end{eqnarray}
The mean values of all $z$-derivatives are zero due to BCs;
additionally,
 $\la T_2\ra=0$.
The remainder is
\begin{equation}
\partial_{t_4}\theta+\nabla_{\!2}\cdot\la\vec{u}_4T_0+\vec{u}_2T_2\ra=0\,.
\label{lstc-407}
\end{equation}
With $T_0$, $T_2$, $p_2$, and $\vec{u}_2$ known, one can find
\begin{eqnarray}
&&\hspace{-8mm}
\textstyle
\nabla_{\!2}\!\cdot\!\la\vec{u}_4T_0\ra
 \!=\!-\!\nabla_{\!2}\!\cdot\!\la T_0\nabla_{\!2}p_2\ra
 \!=\!\frac{2}{21}\Delta_2^2\theta\!
 -\!\!\nabla_{\!2}\!\cdot\!(\theta\nabla_{\!2}\pi_2)\,;
\label{lstc-408}\\[3pt]
&&\hspace{-8mm}
\textstyle
\nabla_{\!2}\!\cdot\!\la\vec{u}_2T_2\ra
 =-\nabla_{\!2}\!\cdot\!\left(\frac{6}{5}\nabla_{\!2}\theta(\nabla_{\!2}\theta)^2-q_2\nabla_{\!2}\theta\right).
\label{lstc-409}
\end{eqnarray}

Substituting (\ref{lstc-408}) and (\ref{lstc-409}) into
Eq.\,(\ref{lstc-407}), one obtains the {\em slow evolution
equation} for $\theta$ in the final form which is similar to
Eq.\,(\ref{lstc-01});
\begin{equation}
\textstyle
\partial_t\theta+\vec{U}\!\cdot\!\nabla\theta+\frac{2}{21}\Delta^2\theta
-\nabla\!\cdot\!\left(\frac{6}{5}\nabla\theta(\nabla\theta)^2-q\nabla\theta\right)=0.
\label{feq}
\end{equation}
Here the imposed advection
$\vec{U}\equiv-\nabla\pi=\la\vec{u}\ra$,
$\nabla\!\cdot\!\vec{U}=0$, subscripts ``2'' for differential
operators are omitted as all the fields depend on $x$ and $y$
only, indexes indicating the smallness order are also omitted as
this equation remains valid in original terms (without formal
smallness parameter $\e$). From
Eqs.\,(\ref{lstc-208},\ref{lstc-209}) the fluid flow up to the
leading order of accuracy is
\[
\vec{v}=6(1-2z)\nabla\theta(x,y)
 +6(z^2-z)\,\Delta\theta(x,y)\vec{e}_z\,.
\]
Advection speed $U\propto\e^3$ is small against $u\propto\e$ when
flow is excited, but is important due to its properties: in
contrast to $\vec{u}$, $\vec{U}$ provides a nonzero gross fluid
flux through a vertical cross-section of the layer.

Note, the heating inhomogeneity makes the quiescent state
impossible, and from Eqs.\,(\ref{lstc-214}) and (\ref{lstc-406})
it follows that below the convective instability threshold, when
$\theta$ decays to zero, the establishing state has nontrivial
\begin{equation}
\textstyle
w_\mathrm{BG}=(-2z^3\!+3z^2\!-\!z)\Delta q,
\quad
T_\mathrm{BG}=\left(\frac{1}{2}\!-\!z\right)q-z.
\label{BGSt}
\end{equation}
However, these fields are small against the fields excited above
the threshold.

Remarkably, Eq.\,(\ref{feq}) holds for the case of a uniform
heating and a weakly inhomogeneous porous matrix. In the most
general case, $q(x,y)$ should be replaced with
\[
 q_\mathrm{gen}(x,y)=q(x,y)+\frac{K(x,y)}{b(x,y)\chi(x,y)}\frac{b_0\chi_0}{K_0}-1
\]
[meanwhile, the background state~(\ref{BGSt}) for this general
case cannot be obtained via the plain replacement of $q$ by
$q_\mathrm{gen}$]. This remark is important for experiments as it
is more convenient to maintain/control a uniform heating of the
layer with a weak inhomogeneity of the structure of the porous
matrix, which is inevitable in real systems. Owing to the same
reason, the physical system we have discussed is the most relevant
one for works~\cite{Goldobin-Shklyaeva} addressing phenomena
related to spatially localized convective currents excited under
parametric disorder [frozen randomly inhomogeneous $q=q(x)$].

DG acknowledges the Foundation ``Perm Hydrodynamics,''
 CRDF (Grant no.\ Y5--P--09--01), and MESRF (Grant no.\ 2.2.2.3.16038)
 for financial support.

\end{document}